\documentclass[prl,twocolumn,groupedaddress]{revtex4b5}
\usepackage{amssymb,amsmath}
\usepackage{latexsym}
\usepackage{graphicx}
\def\scoc{Sr$_2$CuO$_2$Cl$_2$}
\def\chipar{\chi^{(3)}_{xxxx}}
\def\chiperp{\chi^{(3)}_{xxyy}}

\begin{document}

\title{Parity Forbidden Excitations of \scoc\ Revealed by Optical Third-Harmonic
  Spectroscopy}

\author{A.~B.~Schumacher}
\altaffiliation{also Institut f\"ur Angewandte Physik,
  Universit\"at Karlsruhe, 76128 Karlsruhe, Germany}
\author{J.~S.~Dodge}
\altaffiliation[Present address:\  ]{Department of Physics,
  Simon Fraser University, Burnaby, BC~V5A~16S, Canada}
\author{M.~A.~Carnahan}
\author{R.~A.~Kaindl}
\author{D.~S.~Chemla}
\affiliation{Department of Physics, University of California at
   Berkeley, Berkeley, California 94720}
\affiliation{Materials Sciences Division, E.~O.~Lawrence Berkeley
   National Laboratory, Berkeley, California 94720}

\author{L.~L.~Miller}
\affiliation{Ames Laboratory and Department of Physics, Iowa State
   University, Ames, IA 50011}

\date{\today}

\begin{abstract}
  We present the first study of nonlinear optical third harmonic
  generation (THG) in the strongly correlated charge-transfer
  insulator \scoc.  For fundamental excitation in the near-infrared,
  the THG spectrum reveals a strongly resonant response for photon
  energies near 0.7~eV. Polarization analysis reveals this novel
  resonance to be only partially accounted for by three-photon
  excitation to the optical charge-transfer exciton, and indicates
  that an even-parity excitation at 2~eV, with $a_{1g}$ symmetry,
  participates in the third harmonic susceptibility.
\end{abstract}


\maketitle

A central theme of condensed matter physics is the problem of treating
Coulomb interactions on the electron-Volt scale to determine such
low-energy phenomena as the Mott metal-insulator transition,
superexchange, and superconductivity. The two-dimensional copper oxide
plane has emerged as a model system for studying these phenomena.
While excitations at the high-energy scale have been observed in the
the linear optical spectrum of insulating cuprates, the details of
their lineshapes, temperature dependence, and even their assignment
has been controversial~\cite{ref:kastner98}.  In this Letter, we
employ nonlinear optical spectroscopy to determine the symmetry and
spectrum of both even and odd parity states near the charge-transfer
(CT) gap of the model insulating cuprate, \scoc.  Through polarization
analysis of the nonlinear optical response, we find evidence for an
even parity excitation at approximately 2~eV, with $a_{1g}$ symmetry,
and we estimate the magnitude of the dipole matrix element that
couples this excitation to the CT exciton. The technique that we
describe here is generally applicable to correlated insulators, and
should provide valuable input into models of their electronic
structure.

The undoped one-layered oxyhalides of the M$_2$CuO$_2$X$_2$ (M = Ca,
Sr, Ba; X = F, Cl, Br) family are of special interest for studying the
excitation spectrum in cuprates. These compounds share the quasi
two-dimensional CuO$_2$ layers which form the structural and
electronic basis for high-temperature superconductivity. Unlike their
pure oxide counterparts, however, the intervening alkaline earth and
halogen elements provide these materials with a very stable
stoichiometry.  \scoc\ in particular can be obtained in single crystal
form with very high quality~\cite{ref:miller90}. It is regarded as an
almost ideal realization of a two-dimensional, spin-$1/2$ Heisenberg
anti-ferromagnet at half filling and serves as the test compound for
theory in the low-doping regime.

The most obvious optical feature in the insulating cuprates is the CT
gap excitation near 2~eV. Cluster calculations generically predict
additional, optically forbidden CT excitations, with $a_{1g}\ (s)$ or
$b_{1g}\ (d_{x^2-y^2})$ symmetry in
$4/mmm$~\cite{ref:simon96,ref:zhang98,ref:hanamura00}, but presently
there is insufficient experimental guidance for selecting an
appropriate minimal model~\cite{ref:wang96,ref:hasan00,ref:schulzgen01}. Parity
forbidden excitations with ligand field, or $dd$ character, are also
expected; these are intraatomic excitations that result directly from
strong correlations, in which the hole in the ground state ($g$)
Cu($3d_{x^2-y^2}$) orbital is transferred to another Cu orbital of
different symmetry.

Large energy-shift laser Raman spectroscopy provided early evidence
for the existence and energy spectrum of such excitations, with the
assignment of a transition $d_{x^2-y^2}\rightarrow d_{xy}$ at 1.35
eV~\cite{ref:liu93}, also identified in optical
measurements~\cite{ref:perkins93,ref:falck94}. From a ligand-field
analysis of related peaks in the resonant X-ray Raman spectrum, Kuiper
{\em et al.}  have suggested that the $d_{3z^2-r^2}\ (a_{1g})$ state
should be located at 1.5~eV, but were unable to observe it directly
for technical reasons~\cite{ref:kuiper98}.  While a peak in the
$B_{1g}$ laser Raman spectrum of some insulting cuprates may be
consistent with an assignment to this state, this peak is only
observed in $T^\prime$-phase materials~\cite{ref:salamon95}.

In our experiment, we completely characterize the THG third order
susceptibility tensor elements, $\chi^{(3)} (-3\omega; \omega, \omega,
\omega)$, for the Cu--O $\{x,y\}$-plane of \scoc. In the space group
$I4/mmm$ possessed by \scoc\ the nonlinear susceptibility tensor has
two independent elements in the $\{x,y\}$-plane, $\chipar$ and
$\chiperp$, which are in general complex. All tensor elements related
to $\chiperp$\ by permutation of the indices are equal,
$\chi^{(3)}_{yyyy} = \chipar$, and all other tensor elements are zero.
For a driving optical field $E$ at frequency $\omega$, polarized in
the $\{x,y\}$-plane, the nonlinear polarization is given by
\begin{eqnarray}
    \label{eq:polariz}
    P_x(3\omega) & = & \chipar E_x^3 + 3\chiperp E_x E_y^2,\nonumber\\
    P_y(3\omega) & = & \chipar E_y^3 + 3\chiperp E_x^2 E_y.
\end{eqnarray}
Throughout, we take $x$ and $y$ to correspond to coordinates
aligned with the crystalline axes of \scoc, with the $z$-axis
normal to the crystal and in the direction of propagation. It is
useful to write the two tensor elements in terms of an amplitude
and a phase: $\chipar = \kappa_{xx}\:e^{i\alpha}$, and $\chiperp =
\kappa_{xy}\:e^{i\beta}$, with the phase difference defined as
$\delta = \beta - \alpha$.

\begin{figure}[t]
\begin{center}
  \includegraphics[width=0.9\columnwidth]{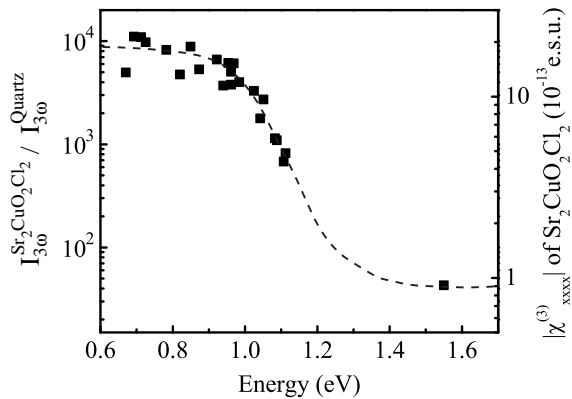}
     \caption{Third harmonic generation in \scoc. The left axis shows the
       relative values of I$(3\omega)\propto|\chipar|^2$. The right
       scale indicates absolute values for \scoc. Dashed line: guide to
       the eye.}
     \label{fig:thg_scocl}
\end{center}
\end{figure}

We first measured the magnitude of the third harmonic susceptibility
$|\chipar|$ at room temperature, as a function of fundamental photon
energy. These and all other measurements described here were performed
with a 250~kHz Ti:sapphire laser amplifier driving an optical
parametric amplifier. To account properly for variations in pulse
duration and mode profile as the laser wavelength is tuned, we
compared the third harmonic intensity I$(3\omega)$ generated in \scoc\ 
with a quartz reference.  We accounted for the \scoc\ absorption at
3$\omega$ in the limit $\ell\ll\lambda$, $\Delta k \ell\approx$0,
appropriate here~\cite{ref:chemla87}. The \scoc\ samples were cleaved
to a thickness of $\ell \sim 100$\ nm, and oriented
crystallographically with X-ray diffraction. The quartz reference was
a thin, $\ell=150\ \mu$m, c-axis plate, whose absolute nonlinear
susceptibility, $\chipar$(quartz)$= 3\times 10^{-14}$~esu, is known
independently~\cite{ref:hermann73}.  The measured relative intensities
along with the resulting $|\chipar|$ of \scoc\ are shown in
Fig.~\ref{fig:thg_scocl}.
The spectrum exhibits a broad $\approx$ 0.5~eV wide resonance at
0.7~eV, varying by more than a factor of 20 over the explored
frequency range. Since the susceptibility
$\chi^{(3)}(-3\omega;\omega,\omega,\omega)$ involves summation over
different intermediate states, the resonance may result from a
parity allowed, three-photon transition at $3\omega = 2.1$ eV, a
parity forbidden, two-photon transition at $2\omega = 1.4$ eV, or
both.  All of these possibilities are consistent with previously
published results~\cite{ref:kastner98}.

\begin{figure}[t]
\begin{center}
     \includegraphics[width=0.9\columnwidth]{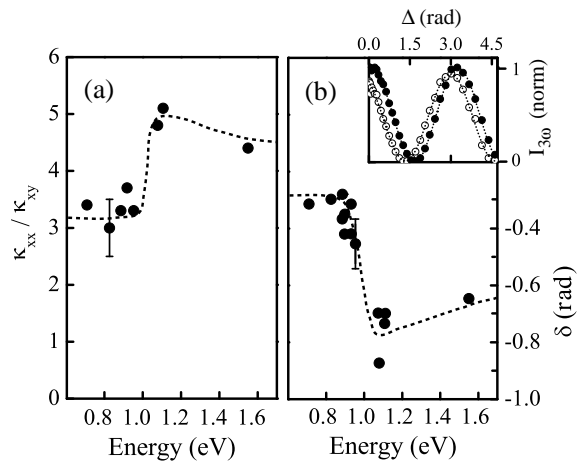}
     \caption{(a) Relative phase $\delta$ and (b)
       relative magnitude $\rho$ of $\chipar$ and $\chiperp$ in \scoc.
       The dashed lines are guides to the eye. Inset: Example of a
       retardance measurement: I$^x(3\omega)$ (filled circles) and
       I$^y(3\omega)$ (open circles) recorded at $\omega =0.89$~eV.
       The dotted lines represent best fits to a $\sin^2$ function,
       from which $\delta$ and $\rho$ are extracted.}
     \label{fig:thg_abph}
\end{center}
\end{figure}

Our chief experimental results are the spectroscopic measurements of
the relative amplitude $\rho = \kappa_{xx}/\kappa_{xy}$ and the
relative phase $\delta$ of $\chipar,\ \chiperp$, shown in
Figs.~\ref{fig:thg_abph}(a) and (b), respectively.
These data were obtained from polarization-sensitive measurements of
the third harmonic optical intensity along each of the two crystalline
axes of the sample, I$^{x,y}(3\omega)$, while varying the polarization
state of the fundamental laser field.  In the experimental geometry
used here, the incident field is initially polarized at 45~degrees to
the two crystalline axes, then passes through a polarization
compensator before entering the sample.  The natural axes of the
compensator are aligned with those of the crystal, so the polarization
state upon entering the sample is given by
$\vec{E}=[\hat{x}E+\hat{y}Ee^{i\Delta}]/\sqrt{2}$, where $\Delta$ is
the compensator retardance. Thus, the nonlinear polarizations $P_x(3\omega)$
and $P_y(3\omega)$ involve controlled mixtures of $\chipar$\ and $\chiperp$,
with $\Delta$ as the control parameter, as may be seen through
Eq.~\ref{eq:polariz}.

As the compensator is adjusted, the THG intensities
I$^x(3\omega)\propto |P_x(3\omega)|^2$ and I$^y(3\omega)\propto |P_y(3\omega)|^2$ both
oscillate sinusoidally with $\Delta$.  Results from a typical
measurement, for $\hbar\omega = 0.89$~eV, are shown in the inset of
Fig.~\ref{fig:thg_abph}. Straightforward manipulation of
Eq.~\ref{eq:polariz} tells us further that the phase offset in
$\Delta$ between these two intensity curves is $\delta$, and that the
minima of I$^{x,y}(3\omega,\Delta)$ are offset from zero by
$\kappa_{xy}(\rho - 3)$. From a series of such measurements at various
laser frequencies, we obtain the dispersion of the relative amplitude
$\rho(\omega)$ and relative phase $\delta (\omega)$, shown in
Figs.~\ref{fig:thg_abph}(a) and (b), respectively.  These data show
that $\rho(\omega)$ and $\delta(\omega)$ both undergo clear,
step--like changes at about 1~eV. Since these are relative quantities,
we can derive the model-independent conclusion that as the photon
energy crosses $\hbar \omega \approx 1$ eV, the transition dipole
elements involved in $\chipar$ and $\chiperp$ connect states with
different symmetry. We show below that this can only be the result of
an excitation with $a_{1g}$ symmetry at 2~eV.

The tensor elements of the nonlinear susceptibility are given by
\begin{equation}
\label{eq:chi3} \chi^{(3)}_{ijkl}(-3\omega;\omega,\omega,\omega) =
\frac{N}{\hbar^3}\mathcal{P}_{I}\sum^{8}_{n = 1}\, {\cal
K}^{(n)}_{ijkl}(-3\omega,\omega,\omega,\omega),
\end{equation}
where the intrinsic permutation operator $\mathcal{P}_I$ applies to
the last three tensor indices $\{jkl\}$~\cite{ref:boyd92}. The eight
terms ${\cal K}^{(n)}$ correspond to different summations over
intermediate states. The dominant term in our experiment is ${\cal
  K}^{(1)}$, which may be written
\begin{equation}
\label{eq:chi3term} {\cal K}^{(1)}_{ijkl} = \sum_{\nu_1,\nu_2,\nu_3}
\frac{\mu^{i}_{g\nu_1}\mu^{j}_{\nu_1\nu_2}\mu^{k}_{\nu_2\nu_3}\mu^{l}_{\nu_3
    g}}{(\Omega_{\nu_1 g}-3\omega)(\Omega_{\nu_2 g}-2\omega)(\Omega_{\nu_3 g}-
    \omega)},
\end{equation}
where $\{\nu_1,\nu_2,\nu_3\} $ are the intermediate states,
$\mu^{i}_{\nu_1\nu_2} = \langle \nu_1|\mu^i|\nu_2\rangle$ are dipole
matrix elements along the direction $\hat{r}_i$, and each
$\Omega_{\nu_\ell g} = \omega_{\nu_\ell g} - i\gamma_{\nu_\ell g}$ is
a complex frequency associated with a transition from the ground state
$g$ to an intermediate state $\nu_j$. Individual terms in
Eq.~\ref{eq:chi3term} may be one-, two-, or three-photon resonant, or
it may contain double or triple resonances, depending on the location
and symmetry of the states $\nu_\ell$. If the product of matrix
elements in Eq.~\ref{eq:chi3term} is to be nonzero, symmetry requires
both that $\nu_1$ and $\nu_3$ transform as $e_u$ in $4/mmm$, and that
$\nu_2$ transforms as $a_{1g}\ (z^2),\ a_{2g}\ (x^3y-xy^3),\ b_{1g}\ 
(x^2-y^2)$ or $b_{2g}\ (xy)$. Moreover, the summation over all
possible permutations in Eq.~\ref{eq:chi3} leads to additional
cancellations, so that those terms in which $\nu_2$ transforms as
$a_{2g}$ or $b_{2g}$ do not contribute to the THG signal.
Consequently, the even-parity states which contribute to the THG
signal are exclusively those with $a_{1g}$ or $b_{1g}$ symmetry.

To evaluate the role of these intermediate states in the measured
third harmonic spectrum, we compare our results from \scoc\ to the
results of a phenomenological, independent-level model, commonly used
in nonlinear optics~\cite{ref:boyd92}.  We assume that the excited
states are independent levels, each with its own energy, dephasing
rate, and symmetry. To obtain a realistic estimate of the matrix
elements, excited state energies, and decay rates of the odd-parity
states, observed in conventional linear optics, we performed a fit of
four lorentzian functions to the experimentally determined absorption
spectrum~\cite{ref:choi99,ref:rein00}. By attributing each of these
lorentzians to a different state $\eta_n$, where $n$ ranges from one
to four, we obtain from the fit the energies $E_n$, dephasing rates
$\gamma_n$, and dipole matrix elements $\mu^x_{gn}$ connecting each
state $\eta_n$ to the ground state. These parameters are listed in
Table~\ref{tab:fitparms}.
\begin{table}[b]
\caption{Energies $E_n$, dephasing rates $\gamma_n$, and matrix
  elements $\mu^x_{gn}$, $\mu^x_{en}$ used in the model calculations
  described in the text.}
\label{tab:fitparms}
\begin{ruledtabular}
\begin{tabular}{lcccc}
$\eta_n$ & $E_n$ & $\gamma_n$ & $\mu^x_{gn}$ & $\mu^x_{en}$\vspace{1pt}\\
 & (eV) & (eV) & $(10^{-18}$\ esu) & $(10^{-18}$\ esu)\\\hline
$\eta_1$ & 1.96 & 0.25 & 1.86 & 3.72\\
$\eta_2$ & 2.44 & 1.75 & 4.00 & 0.80\\
$\eta_3$ & 4.00 & 5.75 & 4.70 & 0.94\\
$\eta_4$ & 5.00 & 5.75 & 4.11 & 0.82
\end{tabular}
\end{ruledtabular}
\end{table}
We emphasize that the parameters used here are chosen merely to yield
a good description of the {\em linear} susceptibility, and
subsequently remain fixed for the nonlinear susceptibility calculation
using Eq.~\ref{eq:chi3}.

\begin{figure}[t]
\begin{center}
     \includegraphics[width=0.9\columnwidth]{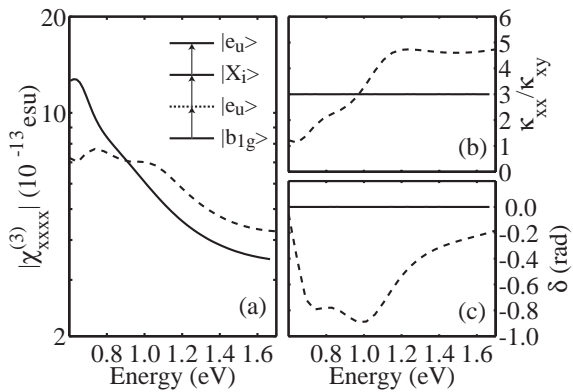}
     \caption{
       Comparison of two model calculations for $\chi^{(3)}$: (a)
       magnitude, (b) relative amplitude $\rho$, and (c) relative
       phase $\delta$.  The inset shows the level scheme assumed in
       the model, with $|X_i\rangle$ the two-photon states of
       interest. In all panels, solid lines indicate that only the
       virtual excitation to $|X_1\rangle = |b_{1g}\rangle$ is
       included.  Dashed lines indicate the addition of a resonant
       state at 2.0~eV, $|X_2\rangle = |a_{1g}\rangle$.}
     \label{fig:model-cal}
\end{center}
\end{figure}

The results of this model calculation are shown as solid lines in
Fig.~\ref{fig:model-cal}.
Panel (a) shows that the absolute magnitude of $\chipar$ exhibits a
peak near 0.7~eV, roughly one third of the energy of the first
odd-parity excited state $\eta_1$ of the model, as expected from a
three-photon resonance. Panels (b) and (c) show the that the relative
amplitude and phase are completely flat as a function of frequency,
with the value of the amplitude ratio $\rho$ fixed at three, while the
phase difference $\delta$\ is fixed at zero. This is the behavior
expected of a spherically symmetric nonlinear
hyperpolarizability~\cite{ref:maker65}.  It should be noted that in
this case, the relative quantities $\rho$ and $\delta$ are entirely
insensitive to the details of the model parameters, because their
values are determined by symmetry. We find that in general, if the
ground state possesses $b_{1g}$ symmetry, then {\em any} set of
excited states containing only $b_{1g}$ and $e_u$ symmetries will
display this overall, spherically symmetric hyperpolarizability, over
the entire frequency spectrum. The model, however, does give the right
order of magnitude for $|\chipar|$. In the case of \scoc, then, the
magnitude of the nonlinear susceptibility is dominated by a
three-photon resonance to CT excited states, and is only weakly
enhanced by two-photon resonances to even-parity
states~\cite{ref:kishida00}. 

While the overall magnitude and shape of the resonance in $\chipar$ is
explained well with only the optically allowed CT excited states, the
structure that we observe in $\delta$ and $\kappa_{xx}/\kappa_{xy}$
implies the involvement of excitations with different symmetries. As
we have seen, the only other states that may contribute to the THG
signal are those with $a_{1g}$ symmetry.  Moreover, when these states
are included in Eq.~\ref{eq:chi3}, the transitions to the $a_{1g}$ and
$b_{1g}$ intermediate states will interfere. Since $\langle
x|x|a_{1g}\rangle = \langle y|y|a_{1g}\rangle$, and $\langle
x|x|b_{1g}\rangle = -\langle y|y|b_{1g}\rangle$, this interference
will exhibit different behavior in $\chipar$ and $\chiperp$, to
produce exactly the deviations from spherical symmetry that we
observe. Thus, we extend our model calculation by adding an excitation
with $a_{1g}$\ symmetry at 2.0~eV.  The additional matrix elements
$\mu^x_{en}$ coupling this excited state to the four odd-parity
excited states are given in Table~\ref{tab:fitparms}. Remarkably, the
results shown as dashed lines in Figs.~\ref{fig:model-cal} (b,c) are
in close agreement with the experiment, both in sign and in magnitude.
There is considerable uncertainty in the parameters obtained from this
simplified model, so the error on our estimate of the energy of
the $a_{1g}$ excitation may be as large as 0.4~eV. However, even with
this error, the midinfrared range is clearly
excluded~\cite{ref:perkins93}, and the symmetry of the state is
unambiguously determined through this analysis.

Finally, we turn to the assignment of this excitation.
Symmetry-allowed alternatives include a CT exciton, a Cu $dd$
excitation, and coupled modes, such as an exciton-phonon (i.e.,
$a_{1g} \in e_u\otimes e_u$). It is likely, however, that {\em all} of
these excitations are active in this range, and are strongly mixed.
This proposal is supported by the anomalously broad Urbach tail
observed at low temperatures in \scoc, which has been explained
recently by a model that involves strong coupling between the
charge-transfer gap excitation and another, optically forbidden,
excitation nearby in energy~\cite{ref:rein00}.  Electronic structure
calculations place both the Cu $dd$ and the $a_{1g}$ CT excitons near
2~eV~\cite{ref:kuiper98,ref:simon96,ref:hanamura00}, and both modes
would be expected to couple strongly to the $e_u$ CT exciton via
phonons. As our theoretical understanding of these excitations grows,
our estimate of the matrix elements $\mu^x_{en}$ may provide a further
test of this scenario.

In summary, we have determined the spectrum of the full THG nonlinear
optical susceptibility tensor for \scoc, using both absolute
measurements against a standard reference and relative measurements
based on polarization analysis. The dominant feature in the absolute
measurement is a peak at $\hbar\omega \sim 0.7$~eV, which can be
explained as a three-photon resonance to the CT gap. The relative
measurements show that an additional, optically forbidden state near
2~eV participates in the THG spectrum, and that this state possesses
$a_{1g}$ symmetry. Both the symmetry and the energy of this state are
consistent with a mixed mode involving both a $d_{x^2-y^2}\rightarrow
d_{3z^2-r^2}$ transition and a coupled CT-exciton-phonon mode. These
measurements were performed with fundamental frequencies well away
from resonance, in contrast to previous Raman work, and the relative
measurements have enabled the estimation of the optical matrix
elements coupling the even-parity state to the CT gap excitation.  In
general, both Raman and THG spectroscopy are different cuts through
the three-dimensional frequency space that $\chi^{(3)}$ may probe.
Further development of nonlinear optical spectroscopy will provide a
more complete determination of these and other optical excitations.

\begin{acknowledgments}
  This work was supported by the U.S.  Department of Energy under
  Contracts No.  DE-AC03-76SF00098 and W-7405-Eng-82. Fellowships from
  the German National Merit Foundation (ABS), and the Deutsche
  Forschungsgemeinschaft (RAK) are gratefully acknowledged, as is the
  assistance of Z.~Rek, Z.~Hussain and X.~Zhou for their assistance in
  orienting the \scoc\ platelet. We thank G.~A.~Sawatzky and
  M.~V.~Klein for their comments on the manuscript.
\end{acknowledgments}

\end{document}